\documentclass[prl,footinbib,twocolumn]{revtex4}
\newcommand{\tnote}[1]{}

\usepackage{hyperref}
\usepackage{amsmath}
\usepackage{amsthm}
\usepackage{amssymb}
\usepackage{graphicx}

\usepackage{guruarticle}
\usepackage{braket}

\newcommand{\authornote}[1]{}

\renewcommand{\vect}[1]{\vec{#1}}
\newcommand{\comment}[1]{}
\renewcommand{\Im}{{\rm Im \;}}
\renewcommand{\expec}[1]{\ensuremath{\left\langle #1 \right\rangle}}

\newcommand{\insertTableOne}{
\begin{table}[t]
\caption{Typical energy scales $E$ and the corresponding time scales $\tau=h/E$ relevant to F\"orster  Resonant Energy Transfer in several systems. $\Delta$ is the F\"orster coupling between the two chromophores, $\epsilon$ is the difference in energy of their first excited states. $\omega_c$ is the high frequency cut-off of the spectral density $J(\omega)$ in the relevant spin-boson model. Typically, $1/\omega_c=\frac{2\epsilon_\infty+\epsilon_p}{2\epsilon_s+\epsilon_p} \tau_D$, where $\tau_D$ is the Debye relaxation time of the solvent and $\epsilon_s$ and $\epsilon_i$ are the solvent's static and high frequency dielectric constants respectively.  LH-I and LH-II are the photosynthetic light harvesting complexes, and ``BChl'' is the bacteria chlorophyll molecule.  We observe that the fluorescence lifetime $\tau_{\mbox{rad}}$ is much longer than the other time scales, suggesting that all other processes of interest occur before radiative decay. We note also that both $\Delta$ and $\epsilon$ span two orders of magnitude, so we might expect very different behaviour for BChl's in LH-II than for typical FRET spectroscopy, such as between green and red fluorescent proteins.}
\begin{center}
\begin{tabular}{|c|c|c|c|c|}
\hline
\multicolumn{2}{|c|}{System} & E (meV) & $\tau$ (ps) & Ref \\
\hline
300K & $k_B T$ & $25$ & 0.16 &  \\
H$_2$O $J(\omega)$ cut-off & $\hbar\omega_c$ & $2-8$ & 0.5-2 &  \cite{Gilmore04} \\ 
THF $J(\omega)$ cut-off  & $\hbar\omega_c$ & $2-4$ & 1-2.5 & \cite{Horng95} \\
\hline
Radiative lifetime & $h/\tau_{\mbox{rad}}$ & $4 \times 10^{-4} $ & $10^4$  & \cite{vanHolde98} \\
Protein relaxation time & $h/\tau_p$ & 0.025 & 162 & \cite{Loffler97} \\
\hline
 Typical FRET  & $\Delta$ & 0.2--2 & 2--20 & \cite{Volker98}\\
(Green $\rightarrow$ red) & $\epsilon$ & 500 & $8 \times 10^{-3}$ &  \\
\hline
BChl in LH-II & $\Delta$ & 46-100 & 0.04 - 0.08& \cite{Hu97JPhysChem} \\
& $\epsilon$ & 0 & & \cite{Hu97JPhysChem} \\
LHII $\rightarrow$ LHII & $\Delta$ & 0.3 & 13 &  \cite{Hu97JPhysChem} \\
& $\epsilon$ & 0 &  & \cite{Hu97JPhysChem} \\
LHII $\rightarrow$ LHI   & $\Delta$ & 0.6 & 7 & \cite{Hu97JPhysChem} \\
& $\epsilon$ & 6.5 & 0.6  & \cite{Hu97JPhysChem} \\
\hline
\end{tabular}
\end{center}
\label{tab:scales}
\end{table}
}

\newcommand{\insertTableTwo}{
\begin{table}[!t]
\caption{Behaviour of $P(t)=\expec{\sigma_z(t)}$, which gives the location of the excitation as a function of time $t$, for $\epsilon=0$ and $\Delta \ll \hbar\omega_c$, where $\Delta$ is the F\"orster coupling strength and  $\omega_c$ is the high frequency cut-off of the spectral density $J(\omega)$.  ``loc'' refers to localisation, ``coh'' to damped coherent oscillations and ``inc'' to incoherent behaviour i.e., exponential decay.  $T$ is the temperature of the system and $\alpha$ is the dimensionless coupling constant of the chromophores to the environment as defined in Equation \protect\eqref{eq:alpha}.   $\tau$ refers to the relaxation rate in the expression $P(t)=\exp(-t/\tau)$.  The analytic form of $P(t)$ is given where known, and is generally valid only for timescales longer than $1/\omega_c$. \mbox{$\Delta_r=\Delta(\Delta/\omega_c)^{\alpha/(1-\alpha)}$}  }
\begin{center}
\begin{tabular}{|c|c|c|c|c|}
\hline
$\alpha$ & $T$ & Key & $P(t)$ & Ref\\
\hline
$\alpha>1$ & $T=0$ & loc & $P(t)=1$, all $t$ & \cite{Leggett87}\\
$0<1/2$ & T=0 & coh & \begin{minipage}{4.2cm}
\mbox{$P(t) \approx \exp\left[ -2 t \frac{\Delta_r}{\pi\alpha} \sin^2 \frac{\pi \alpha}{2(1-\alpha)} \right]$}\\
$\times \cos \left[ t \frac{\Delta_r}{\pi\alpha} \sin \frac{\pi \alpha}{1-\alpha} \right]$
\end{minipage} & \cite{Lesage98} \\
$0<1/2$ & $T > T^\ast$ & inc & $ \tau^{-1} = \frac{\Delta^2}{\omega_c} \frac{\sqrt{\pi}}{2} \frac{\Gamma(\alpha)}{\Gamma(\alpha+1/2)} \left[ \frac{\pi k T}{\hbar\omega_c} \right]^{2\alpha-1} $ &\cite{Leggett87} \\
$> 1/2$ & $T>0$ & inc & \texttt{"} & \cite{Leggett87}\\
\hline
\end{tabular}
\end{center}
\label{tab:alpha behaviour}
\end{table}
}

\begin{document}
\title{Criteria for quantum coherent transfer of excitons between chromophores in a polar solvent}
\author{Joel Gilmore}
\author{Ross H. McKenzie}
\affiliation{Department of Physics, University of Queensland, Brisbane, Qld 4072, Australia}
\date{\today}
\begin{abstract}
We show that the quantum decoherence of F\"orster resonant energy transfer between two optically active molecules can be described by a spin-boson model. This allows us to give quantitative criteria, in terms of experimentally measurable system parameters, that are necessary for coherent Bloch oscillations of excitons  between the chromophores. Experimental tests of our results should be possible with Flourescent Resonant Energy Transfer (FRET) spectroscopy. Although we focus on the case of
protein-pigment complexes our results are also relevant to quantum dots and organic molecules
in a dielectric medium. 

\end{abstract}
\maketitle

Decoherence is the process whereby quantum interference effects are ``washed out'' by the interaction of a quantum system with its environment. It has been suggested that decoherence is responsible for the crossover from quantum to classical behavior \cite{Zurek03RevModPhys}. Decoherence places limits on the possibility of quantum computation \cite{DiVincenzo00}. The challenge of building a quantum computer and the potential of biomimetics \cite{Sarikaya03} raises the possibility of exploiting the self assembly of complex biomolecular systems, such as light harvesting photosynthetic protein-pigment complexes \cite{Lovett03PRB}. However, this also raises profound questions, which are of interest in their own right, about what role quantum effects play in biomolecular functionality. One model for describing quantum decoherence is the spin-boson model \cite{Leggett87,Weiss99}. It describes quantum tunneling between two quantum states that are coupled to a dissipative bath which is modelled by a set of harmonic oscillators (see the Hamiltonian in eqn. \eqref{eq:spin boson standard} below). This model has been used to describe systems ranging from Josephson junction qubits \cite{Makhlin01} to electron transfer in biomolecules \cite{Renger02,Xu94ChemPhys}. In this Letter, we show how the spin-boson model can also be used to describe the transfer of excitons between two chromophores by the mechanism of F\"orster resonant energy transfer in a dielectric medium. Established results for the spin boson model are then used to give stringent criteria for quantum coherent exciton transfer. Our results are also relevant to quantum dots (see for example \cite{Jones03}) and small molecules in a dielectric medium \cite{Medintz03}.

\begin{figure}[t]
\begin{center}
\includegraphics[width=6cm]{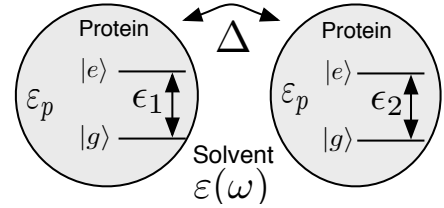}
\caption{Two chromophores with energy gaps $\epsilon_1$ and $\epsilon_2$ are coupled with an interaction energy $\Delta$ due to the F\"orster dipole-dipole interaction.  In the simplest model, the chromophores are centred inside spherical proteins with static dielectric constant $\epsilon_p$, and surrounded by a polar solvent with frequency dependent dielectric constant $\epsilon(\omega)$.}
\label{fig:transfer}
\end{center}
\end{figure}

{\it Model for interaction of the individual chromophores with the solvent}.  In can be shown \cite{Gilmore04} that the coupling of the electronic excitations in a chromophore to its environment may be modelled by an independent boson model \cite{Mahan90} of the form
\begin{equation}
H = \frac{1}{2} \epsilon \sigma_z + \sum_\alpha \omega_\alpha a_\alpha^\dagger a_\alpha + \frac{1}{2}( \Delta\mu) \; \sigma_z \;\op{R}.
\end{equation}
Here the chromophore is treated as a two level system with energy gap $\epsilon$ between the ground and excited state, and $\Delta\mu$ is the difference between the dipole moment of the chromophore in the ground and excited states.  $\op{R}=\sum_\alpha C_\alpha (a_\alpha+a_\alpha^\dagger)$ is the quantised reaction field \cite{Onsager36,Bottcher73} experienced by the chromophore dipole due to the ``cage'' of polarised solvent and protein  molecules around it.  The $C_\alpha$ are the couplings of the excitation to each mode. The coupling to the environment and quantum dynamics is completely specified by the spectral density,
\begin{equation}\label{eq:J(omega) defn}
J(\omega)= 4\pi (\Delta\mu)^2 \sum_\alpha C_\alpha^2 \delta(\omega-\omega_\alpha).
\end{equation}

In the simplest picture of protein-pigment complexes, the chromophore can be treated as a point dipole inside a uniform, spherical protein \cite{Hofinger01,Bottcher73} surrounded by a uniform polar solvent \cite{Onsager36}.  Typical dielectric relaxation times of proteins \cite{Loffler97} are significantly longer than the other time relevant scales (except fluorescence lifetimes, see Table \ref{tab:scales}) and so we consider only the static dielectric constant for the protein.  The spectral density \cite{Gilmore04} is then
\begin{equation}
J(\omega) =  \frac{(\Delta\mu)^2}{4\pi\epsilon_0 b^3} \Im \frac{2(\epsilon(\omega)-\epsilon_p)}{2\epsilon(\omega)+\epsilon_p}.
\end{equation}
where $b$ is the radius of the protein containing the chromophore, $\epsilon(\omega)$ is the complex dielectric function of the solvent and $\epsilon_p$ is the static dielectric constant of the protein.  For the case of a Debye solvent \cite{Hsu97},
\begin{equation}\label{eq:J(omega) final}
J(\omega) = \frac{(\Delta\mu)^2}{2\pi\epsilon_0 b^3} \frac{6\epsilon_p (\epsilon_s-\epsilon_\infty)}{(2\epsilon_s+\epsilon_p)(2\epsilon_\infty+\epsilon_p)} \frac{\omega\tau_E}{\omega^2\tau_E^2+1},
\end{equation}
where $\epsilon_s$ and $\epsilon_\infty$ are the static and high frequency frequency dielectric constants of the solvent respectively, and $\tau_E = \frac{2\epsilon_\infty+\epsilon_p}{2\epsilon_s+\epsilon_p} \tau_D$ where $\tau_D$ is the Debye relaxation time of the solvent.  For water at room temperature, these parameters are $\epsilon_s=78.3$, $\epsilon_\infty=4.21$ and $\tau_D = 8.2ps$ \cite{Afsar78} while for THF (tetrahydrofuran) they are $\epsilon_s=8.08$, $\epsilon_i=2.18$ and $\tau_D=3ps$ \cite{Horng95}.  Typical protein static dielectric constants $\epsilon_p$ are between 4--40 depending on which part of the protein is of interest \cite{Pitera01,Hofinger01,Loffler97}.   $\tau_E$ therefore takes values between $0.5-2.5ps$.

{\it Model for FRET in the presence of a solvent}. We now consider the case of two biomolecules coupled by the F\"orster interaction \cite{Forster65}.  This is a dipole-dipole interaction which produces a non-radiative transfer of an excitation between two chromophores (see Figure \ref{fig:transfer}).  This interaction is the basis for energy transportin photosynthetic light harvesting complexes and fluorescent resonant energy transfer (FRET) spectroscopy. We may write the total Hamiltonian as the sum of two spin-boson Hamiltonians for each chromophore \cite{Gilmore04}:
\begin{align}\label{eq:fret}
H = &  \frac{1}{2} \epsilon_1\sigma_z^1 + \frac{1}{2} \epsilon_2 \sigma_z^2 +(\Delta \mu_1) \sigma_z^1 \op{R}_1  +  (\Delta \mu_2) \sigma_z^2 \op{R}_2 + \; \nonumber\\
& \Delta (\sigma_x^1\sigma_x^2 + \sigma_y^1 \sigma_y^2) + \op{B}_1 +  \op{B}_2,
\end{align}
where $\op{R}_i =\sum_\alpha C_{i,\alpha} (a_{i,\alpha} + a_{i,\alpha}^\dagger)$, ($i=1,2$) is the quantised reaction field operator for molecule $i$, and $\op{B}_i = \sum_\alpha \omega_{i,\alpha} a_{i,\alpha}^\dagger a_{i,\alpha}$ is the energy stored in the solvent cage of molecule $i$.  For molecules sufficiently far apart ($\geq 20$\AA), the interaction \cite{Forster65} is given by
\begin{equation}
\Delta=\frac{\kappa \mu_1\mu_2}{n^2 R^3}
\end{equation}
where $\mu_i=\Braket{e|\op{\mu}|g}_i$, $i=1,2$, is the transition dipole moments of the chromophores (distinct from the change in dipole moment of the molecule during the transition, $\Delta\mu_i$), $n$ is the refractive index of the solvent, $R$ is the separation of the molecules and $\kappa$ is related to the relative orientation of the two dipoles \cite{DipoleNote}.  

For future convenience, in matrix notation the Hamiltonian \eqref{eq:fret}
\begin{eqnarray}\label{eq:H matrix}
H&=&  \sum_{
\begin{minipage}{20pt}
\center
\tiny $\alpha$
$i=1,2$
\end{minipage}} \omega_{i,\alpha} a_{i,\alpha}^\dagger a_{i,\alpha}+ \\
&&\left(\begin{array}{cccc}

\epsilon_{+} + \op{V}_{+}		&	0	&	0		&	0		\\
0		&	\epsilon_{-} + \op{V}_{-}		&	\Delta		&	0		\\
0		&	\Delta		&	-(\epsilon_{-} + \op{V}_{-})		&	0		\\
0		&	0		&	0	&	-(\epsilon_{+} + \op{V}_{+})
\end{array}
\right),\nonumber
\end{eqnarray}
where
\begin{align}
\op{V}_{\pm} &= \Delta\mu_1 \op{R}_1 \pm \Delta\mu_2 \op{R}_2 \nonumber \\
&= \left[  \Delta\mu_1 \sum_{\alpha} C_{\alpha} (a_{\alpha}+a_{\alpha}^\dagger) \pm \Delta\mu_2 \sum_\beta D_\beta (b_\beta + b_\beta^\dagger) \right],
\end{align}
and $\epsilon_{\pm} =  \epsilon_1\pm \epsilon_2$.

\insertTableOne

{\it Mapping to the spin boson model}.  The number of excitations in the system is related to $\op{N}=\frac{1}{2}(\sigma_z^1+\sigma_z^2+2)$ which we note commutes with the Hamiltonian: $[H, \op{N}]=0$. Hence the total number of excitations is a constant of the motion. Note we are assuming that the fluorescence lifetime of the chromophores is much longer than the other time scales of the system described by $H$ and so do not need to include radiative decay in $H$ (typically $1-10ns$, Table \ref{tab:scales}).  If we consider only singly excited systems then $\op{N}\ket{\Psi} = \ket{\Psi}$, and we can project onto the corresponding two dimensional subspace $\{ \ket{e}_1 \otimes \ket{g}_2,\ket{g}_1 \otimes \ket{e}_2 \}$, where $\ket{g}_i,\ket{e}_i$ represents the ground and excited states respectively of chromophore $i=1,2$.  We note that this is a decoherence-free subspace \cite{DiVincenzo00} with respect to decoherence due to the environment.  We can thus restrict our Hamiltonian to the central $2\times 2$ submatrix of \eqref{eq:H matrix}, which can be written in terms of new Pauli sigma matrices as the spin-boson model \cite{Leggett87}
\begin{equation}\label{eq:spin boson final}
H = \frac{1}{2} \epsilon \sigma_z + \Delta \sigma_x + \sigma_z \cdot \op{V} + \sum_{
\begin{minipage}{20pt}
\center
\tiny $\alpha$
$i=1,2$
\end{minipage}} \omega_{i,\alpha} a_{i,\alpha}^\dagger a_{i,\alpha},
\end{equation}
where $\op{V}\equiv \op{V}_{-}$ represents the interaction with the environment, and $\epsilon\equiv \epsilon_{-}=\epsilon_1-\epsilon_2$ is the difference in energy for the excitation on the different chromophores.   We assume the bath modes coupled to each chromophore are independent, i.e., $[a_{1,\alpha},a_{2,\beta}^\dagger]=0$.  This can be justified if the molecules are sufficiently far apart and their cavities can be treated independently \cite{Jang02}.  The environment can then again be modelled as a set of independent harmonic oscillators \cite{Leggett87} in the standard form of the spin-boson model:
\begin{equation}\label{eq:spin boson standard}
H = \frac{1}{2}\epsilon\sigma_z + \Delta\sigma_x + \sigma_z \sum_\alpha C_\alpha (a_\alpha^\dagger+a_\alpha) + \sum_\alpha \omega_\alpha a_\alpha^\dagger a_\alpha
\end{equation}
where the $a_\alpha$ now include both sets of independent harmonic oscillators.

To complete the description, we must specify the new spectral density $J(\omega)$ which now describe the environments around both molecules jointly.  This may be obtained as for the single molecule case \cite{Gilmore04} by following the ansatz of Caldeira and Leggett \cite{Caldeira83} by treating the fluctuations in the environment via the correlation function $\expec{V(t)V(0)}$:
\begin{align*}
&\expec{V(t)V(0)} =\\
& (\Delta \mu_1)^2 \expec{\op{R}_1(t)\op{R}_1(0)} + (\Delta\mu_2)^2 \expec{\op{R}_2(t)\op{R}_2(0)} + \\
&\Delta\mu_1\Delta\mu_2 \left( \expec{\op{R}_1(t)\op{R}_2(0)} +  \expec{\op{R}_2(t)\op{R}_1(0)} \right).
\end{align*}
Provided that the chromophores are sufficiently far apart that their cages are uncorrelated, $\expec{R_1(t)R_2(0)}=\expec{R_2(t)R_1(0)} = 0$.  $J(\omega)$ is then given by
\begin{align}
J(\omega) &= \Im \int dt e^{i\omega t} \left[i\expec{V(t)V(0)}\theta(t) \right] \\
&= J_1(\omega) + J_2(\omega),
\end{align}
i.e., the new spectral density is simply the sum of the appropriate spectral densities for the individual chromophores.  

Modelling of FRET systems by spin-boson like models has been considered previously \cite{Rackovsky73,Soules71,Jang02}, but only with perturbation theory and Fermi's golden rule. Here we have presented a specific microscopic derivation of the effect of the environment on transfer, and given an explicit form for the spectral density $J(\omega)$ that can be determined from experiments on single chromophores.

We note that we have Ohmic dissipation, i.e., $J(\omega)=\alpha \omega$ for $\omega<\omega_c=h/\tau_E$ with the dimensionless coupling constant
\begin{equation}\label{eq:alpha}
\alpha = \frac{1}{2\pi\epsilon_0 h} \left[ \frac{(\Delta\mu_1)^2}{b_1^3} + \frac{(\Delta\mu_2)^2}{b_2^3} \right] \frac{6\epsilon_p (\epsilon_s-\epsilon_\infty)\tau_D}{(2\epsilon_s+\epsilon_p)^2}  .
\end{equation}
If $\epsilon,\Delta \ll \omega_c$, for Ohmic dissipation $\alpha$ is a critical parameter for determining the quantum dynamics \cite{Leggett87,Weiss99,Lesage98} (see Table \ref{tab:alpha behaviour}).  For typical free chromophores in water at room temperature, $\alpha$ of the order $0.1-1$ \cite{Gilmore04}, which represents strong coupling to the environment (in comparison, $\alpha$ is orders of magnitude smaller for Josephson Junction qubits \cite{Makhlin01}.)  However, the protein environment pushes the solvent away from the chromophore, and can make $\alpha$ much less than one.

\begin{figure}[t]
\begin{center}
\includegraphics[width=7cm]{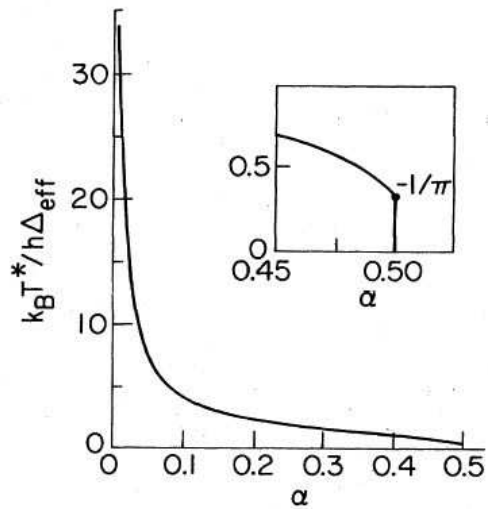}
\caption{Dependence on the environmental coupling $\alpha$ of the critical temperature $T^\ast$ for the cross-over from coherent to incoherent behaviour.  Above the sharp cut-off at $\alpha=1/2$, coherent oscillations do no occur for any temperature.  Reproduced from ref. \cite{Leggett87}.}
\label{fig:alpha vs T}
\end{center}
\end{figure}

{\it Criteria for quantum coherent RET.}  The location of the excitation is given by $P(t)=\expec{\sigma_z(t)}$. Suppose that the excitation is initially localised on one chromophore (the ``donor''), corresponding to $P(0)=+1$, with the other chromophore (the ``acceptor'') in the ground state.   The F\"orster coupling will cause transfer of the excitation between the chromophores.  We now establish the conditions of validity of F\"orster's equation for FRET efficiency in terms of the convolution of the absorption and emission spectras of the two chromophores \cite{Forster65,Jang02} such as is widely used in ``spectroscopic ruler'' applications \cite{Ha01} in molecular biophysics. These results are based on two assumptions (i) second order perturbation theory in $\Delta$ (the Fermi Golden Rule), which assumes that $\Delta$ is small compared to the other energy scales of the system, and (ii) that there is no back transfer, i.e., incoherent transfer of the excitation.  In typical FRET systems, such as between red and green fluorescent proteins, $\Delta$ is typically less than 1meV, while $\epsilon \sim$ 500meV and $\hbar\omega_c \sim$ 8meV (see Table \ref{tab:scales}) which justifies F\"orster's assumption (i). However, assumption (ii) is only justified by Leggett et al.'s nontrivial derivation: provided that $\Delta<\hbar\omega_c$, then for a sufficiently large difference in energy levels  ($\sim \epsilon > \Delta_r=\Delta(\Delta/\omega_c)^{\alpha/(1-\alpha)}$, where $\Delta_r$ is the oscillation frequency renormalised by interaction with the bath environment) or for sufficiently high temperature ($T>T^\ast\sim \Delta_r/\alpha$, see Table \ref{tab:scales} and Figure \ref{fig:alpha vs T}) the transfer is always incoherent \cite{Leggett87}.  We have therefore justified the use of F\"orster's equation for typical FRET systems.

However, when $\epsilon$ is small and $0<\alpha<1/2$ and $T<T^\ast(\alpha)$ (Table \ref{tab:alpha behaviour}) coherent oscillations may occur \cite{Leggett87,Lesage98} and the Fermi Golden Rule derivation for FRET no longer applies.  Further, from Table \ref{tab:scales} we see that $\Delta$ may be comparable to or greater than the reorganisation energy $\hbar\omega_c\alpha$ \cite{Mahan90,Leggett87}.  Therefore,  second-order perturbation theory in $\Delta$ may be hard to justify.  From Table \ref{tab:scales} we see that $\Delta$ takes on a wide range of values, and it is conceivable that at low temperatures and between chromophores with very close (or identical) energy levels that the assumption of incoherent transfer, and hence the Fermi Golden Rule result, may break down.

{\it Experimental tests}. A possible way to observe the coherent oscillations of excitons between chromophores is to use identical chromophores which are at an angle to each other so that their dipole moments are not parallel. Then from a bulk sample the oscillations should be present in the time dependence of both the fluorescent anisotropy and the flourescent noise \cite{Yamazaki02}. This effect has recently been reported for pairs of chromophores (e.g., anthracene dimers) that are covalently bonded and in a solvent at room temperature \cite{Yamazaki02, Sato03}. However, although in all cases the conditions of the experiment are in the regime $\Delta < \omega_c$ they do not satisfy two of the necessary conditions for coherent oscillations, $T < T^\ast \sim \Delta/\alpha$ and $\alpha < 1/2$. For example, reference \cite{Yamazaki02} concerns a molecule DTA in THF (tetrahydrofuran) solvent at room temperature. The measured oscillation period is 1 picosecond and the damping time is also 1 picosecond. For this solvent the cut off frequency is $\hbar \omega_c \simeq 4 $ meV (see Table \ref{tab:scales}), and we expect $\Delta < \hbar \omega_c$ and so the results for Ohmic dissipation should be relevant at the qualitative level.  To determine $\alpha$, we can estimate the reorganisation energy, $E_r \sim \alpha \hbar\omega_c$, from the width of the absorption and fluorescent spectra or from the Stokes shift \cite{Reynolds96}. Roughly, $E_r \sim 100$ meV, and so $ \alpha \sim 25$ and no coherent oscillations should occur.  Furthermore, Monte Carlo simulations that take into account a broad range of $\Delta/\omega_c$ values do not give coherent oscillations for these kind of parameter values (compare Fig. 13 in ref. \cite{Muhlbacher03} and Fig. 7 in ref. \cite{Volker98}).  

{\it Application to photosynthesis}.  One system of particular interest is the transfer of excitations between bacterial light harvesting (LH) complexes (I and II) in photosynthetic units.  In a typical process, an LH-II ring of chlorophyll chromophores absorbs a photon.  The excitation may then transferred to other LH-II rings before reaching the LH-I ring where it is sent to the reaction centre to be converted to chemical energy.  As the chromophores within the ring are identical, $\epsilon\approx 0$ while $\Delta$ is between $46-100$meV.  Here $\Delta \gg \epsilon$, but also $\Delta > \omega_c$, and so the results of \cite{Leggett87} are not applicable. We expect coherent transfer of the excitation around the ring \cite{Carmeli88}.  Experimental studies suggest that the excitation is indeed delocalised \cite{Hu97JPhysChem}.  For transfer between identical LH-II rings, again $\epsilon=0$ but $\Delta \ll \hbar\omega_c$ (see Table \ref{tab:scales}).  In this case, $T>T^\ast(\alpha)$ (Figure \ref{fig:alpha vs T}) and the transfer will be incoherent. Finally, for transfer on an excitation from an LH-II to an LH-I ring, $\epsilon \gg \Delta$ and we expect incoherent transfer.  These behaviours play an important functional role - the delocalisation of the excitation prevents radiative loss of the excitation \cite{Hu97JPhysChem}, while the incoherent inter-ring transport establishes a one-way flow of energy towards the reaction centre.

\insertTableTwo

In conclusion, we have shown that the decoherence of two chromophores coupled by the F\"orster interaction in the presence of a solvent can be described by the spin-boson model. We find that while the standard FRET efficiency formulas do hold for most typical systems, they are based on nontrivial assumptions which may not be true in general.  We give the quantitative conditions necessary for coherent Bloch oscillations of excitons between the chromophores and suggest how this could be tested experimentally.  Finally, we have used these models to describe the transfer of excitations between light harvesting complexes in bacteria and their relevance to the system's biological functionality.

\begin{acknowledgments}
This work was supported by the Australian Research Council and the University of Queensland Graduate School Research Travel Award.  We thank P. Burn, I. Samuel, A. Doherty, T. Simonson, B. Lovett, A. Nazir and G. Milburn, for helpful discussions.  We thank A. D. Briggs and the QIPIR at Oxford for hospitality.
\end{acknowledgments}

\bibliographystyle{apsrev}
\bibliography{phd}

\begin{thebibliography}{37}
\bibitem[{Dip()}]{DipoleNote}
\bibinfo{note}{For dipoles that undergo sufficiently fast Brownian rotation
  \cite{Forster65}, we can use the average value $\expec{\kappa^2}=2/3$. For
  fixed molecules, however, a different value may be required. The given
  expression for the coupling energy is the first term of a multipole
  expansion; for close molecules (such as intra-ring coupling in light
  harvesting complexes in photosynthesis) higher order terms may be required
  \cite{Hu97JPhysChem}.}
\end{thebibliography}

\end{document}